\documentclass[fleqn,10pt]{wlscirep}
\title{High-energy spin fluctuation \\in low-$T_{\rm c}$ iron-based superconductor LaFePO$_{0.9}$}

\author[1,2,3,*]{Motoyuki Ishikado}
\author[2,+]{Shin-ichi Shamoto}
\author[4]{Katsuaki Kodama}
\author[5]{Ryoichi Kajimoto}
\author[5]{Mitsutaka Nakamura}
\author[6]{Tao Hong}
\author[7]{Hannu Mutka}

\affil[1]{Neutron Science and Technology Center, Comprehensive Research Organization for Science and Society (CROSS), Tokai, Naka, Ibaraki 319-1106, Japan}
\affil[2]{Advanced Science Research Center, Japan Atomic Energy Agency (JAEA), Tokai, Ibaraki 319-1195, Japan}
\affil[3]{Electronics and Photonics Research Institute, National Institute of Advanced Industrial Science and Technology (AIST), Tsukuba, Ibaraki 305-8562, Japan}
\affil[4]{Materials Sciences Research Center, Japan Atomic Energy Agency (JAEA), Tokai, Ibaraki 319-1195, Japan}
\affil[5]{J-PARC Center, Japan Atomic Energy Agency (JAEA), Tokai, Ibaraki 319-1195, Japan}
\affil[6]{Neutron Scattering Division, Oak Ridge National Laboratory (ORNL), Oak Ridge, Tennessee 37831, USA}
\affil[7]{Institut Laue-Langevin (ILL), 71 avenue des Martyrs, CS 20156, F - 38042 Grenoble Cedex 9, France}

\affil[*]{m\_ishikado@cross.or.jp}
\affil[+]{shamoto.shinichi@jaea.go.jp}

\begin{abstract}
Spin fluctuations are widely believed to play an important role in the superconducting mechanisms of unconventional high-temperature superconductors. Spin fluctuations have been observed including iron-based superconductors. However, in some iron-based superconductors such as LaFePO$_{0.9}$, they have not been observed by inelastic neutron scattering (INS). LaFePO$_{0.9}$ is an iron-based superconductor with a low superconducting transition temperature ($T_{\rm c}$= 5 K), where line nodes are observed in the superconducting gap function. The line-node symmetry typically originates from sign reversal of the order parameter in spin-fluctuation-mediated superconductivity. This contradiction has been a long-standing mystery of this superconductor. Herein, spin fluctuations were found at high energies such as 30$-$50 meV with comparable intensities to an optimally doped LaFeAs(O,F). Based on this finding, the line-node symmetry can be explained naturally as spin-fluctuation-mediated superconductivity.

\end{abstract}
\begin{document}

\flushbottom
\maketitle
%
\section*{Introduction}

In iron-based superconductors, superconductivity appears in the vicinity of an antiferromagnetic (AF) phase and a structural phase transition from tetragonal to orthorhombic phases. Therefore, spin and multi-orbital dynamics are believed to play an important role in the superconducting mechanisms \cite{Hosono15, Basov, Yamakawa, Yanagi}. The spin dynamics have been studied intensively by inelastic neutron scattering (INS) \cite{Dai}. In the superconducting states, magnetic resonance modes have been observed in the INS spectra of iron-based superconductors \cite{Christianson, Lumsden, Chi, Li}. Based on the doping dependence, the magnetic resonance energies are closely correlated with the superconducting gap energies \cite{Paglione, J.T.Park, Shamoto10, Lee}. The low-energy spin dynamics are well explained by the Fermi surface nesting model\cite{Mazin, Kuroki}. On the other hand, neither the magnetic resonance mode nor the spin fluctuation itself have been observed in the low superconducting transition temperature ($T_c$) iron-based superconductor LaFePO$_{1-y}$ \cite{Taylor}. LaFePO$_{1-y}$ is the first superconductor discovered among the iron-based pnictogen compounds \cite{Kamihara}. Low-energy spin dynamics have also been studied by nuclear magnetic resonance, suggesting that there are no AF spin fluctuations \cite{Nakai}. The spin fluctuations are also strongly suppressed on highly-doped LaFeAs(O,F) and Ba(Fe,Co)$_2$As$_2$ \cite{Wakimoto, Matan}. The suppression of low-energy spin fluctuations in both LaFeAs(O,F) and Ba(Fe,Co)$_2$As$_2$ is explained by the poor nesting condition between Fermi surfaces (FSs) at $\Gamma$ and M points caused by the disappearance of hole FSs with increased electron doping \cite{Kuroki}. After the disappearance of hole FSs, the spin fluctuation energies are expected to increase due to the necessary energy for electron-hole excitation between the two bands \cite{Shamoto}. A theoretical calculation based on a combination of density functional theory (DFT) and dynamical mean field theory (DMFT) successfully reproduced the effective band width of magnetic excitations of NaFeAs in a wide energy range \cite{Zhang}. According to theoretical calculation, spin fluctuations are expected to exist only above 30 meV for LaFePO \cite{Yin}. The pnictogen height ($h_{Pn}$) of LaFePO is relatively low (1.14 ${\rm \AA}$) compared to that of BaFe$_2$As$_2$ (1.37 ${\rm \AA}$), resulting in a wide band width of the magnetic excitations in LaFePO. This is expected to result in a weak INS intensity at high energies \cite{Yin}. 

The parent stoichiometric material LaFePO$_{1.0}$ exhibits no superconductivity, long range magnetic order, or structural phase transition \cite{McQueen}. Nevertheless, electron-doped LaFePO$_{1-y}$ shows line-node symmetry revealed by the temperature dependence of the magnetic penetration depth and thermal conductivity measurements \cite{Fletcher, Yamashita}. The line-node symmetry of the sign-reversing order parameter reminds us of the intervention of the magnetic fluctuation. Therefore, it is intriguing that there is no magnetic order in the parent material of LaFePO$_{1.0}$ in contrast to LaFeAsO$_{1.0}$. Consequently, it is quite important to obtain dynamical information on the spin fluctuations of LaFePO$_{0.9}$ in a wide energy range to examine the effects of the spin fluctuation on the superconductivity with line-node gap symmetry.

\begin{figure}[ht]
\centering
\includegraphics[width=\linewidth]{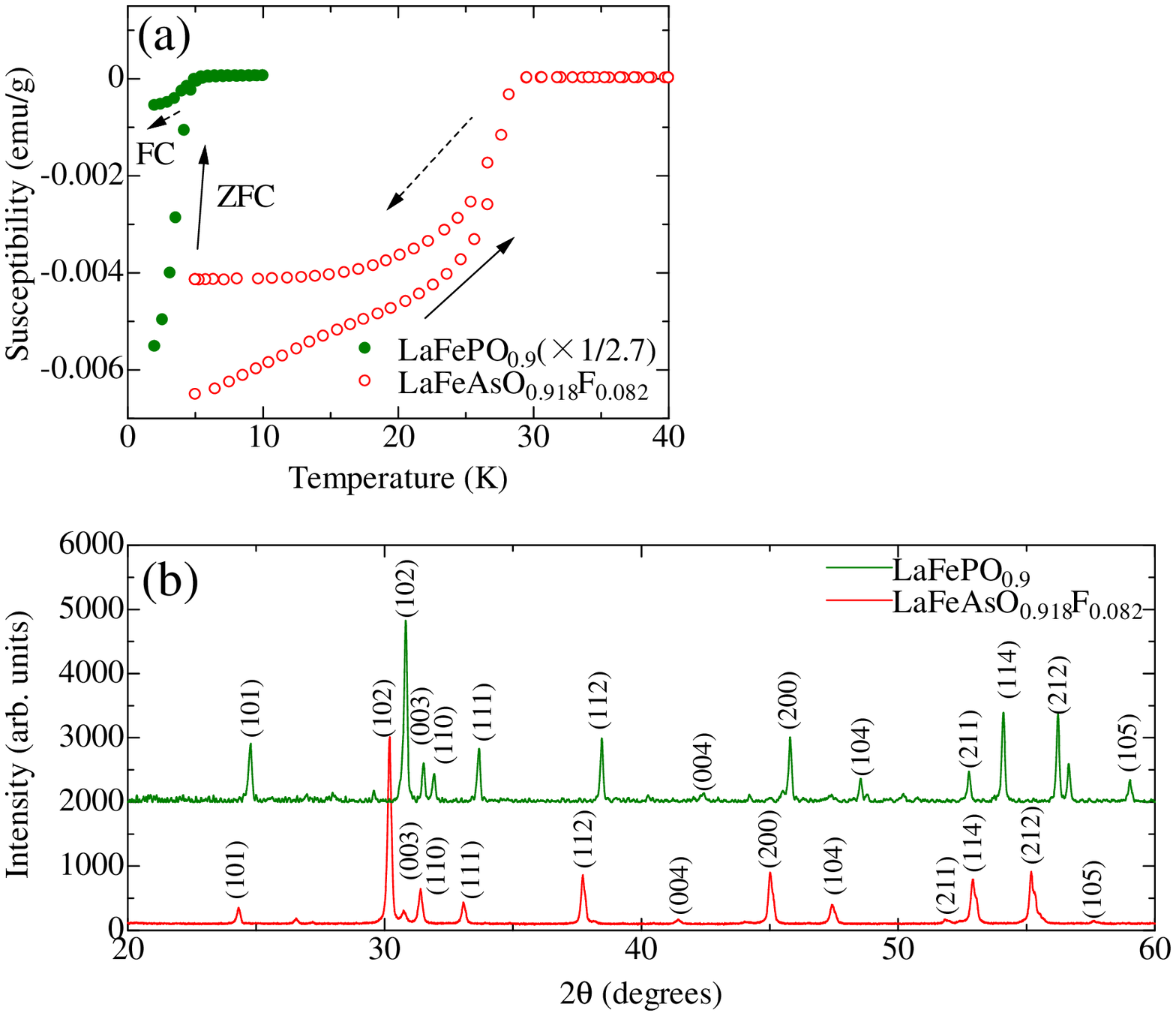}
\caption{Magnetic susceptibilities (a) and X-ray diffraction patterns (b) of measured powder samples of LaFePO$_{0.9}$ and LaFeAsO$_{0.918}$F$_{0.082}$. Zero field cooling (ZFC) and field cooling (FC) processes are shown in (a) by arrows with solid and broken lines, respectively. The diffraction pattern of LaFePO$_{0.9}$ is vertically shifted for clarity.}
\label{Figure1}
\end{figure}

Inelastic neutron scattering measurements have been performed on powder samples of LaFePO$_{0.9}$ with $T_c$= 5 K and optimally doped LaFeAsO$_{0.918}$F$_{0.082}$ (LaFeAsOF) with $T_c$= 29 K as a reference. They were characterized by magnetic susceptibility and X-ray diffraction measurements as shown in Fig. \ref{Figure1}. In this paper, we demonstrate that  spin fluctuation has been clearly observed in low-$T_c$ LaFePO$_{0.9}$ ($T_c$= 5 K) in the range of 30$-$50 meV with similar intensities to the  optimally doped LaFeAsO$_{0.918}$F$_{0.082}$ at the normal state (non-superconducting state), suggesting the universality of the correlation between line-node symmetry and spin fluctuations.

\section*{Results}

\begin{figure}[ht]
\centering
\includegraphics[width=\linewidth]{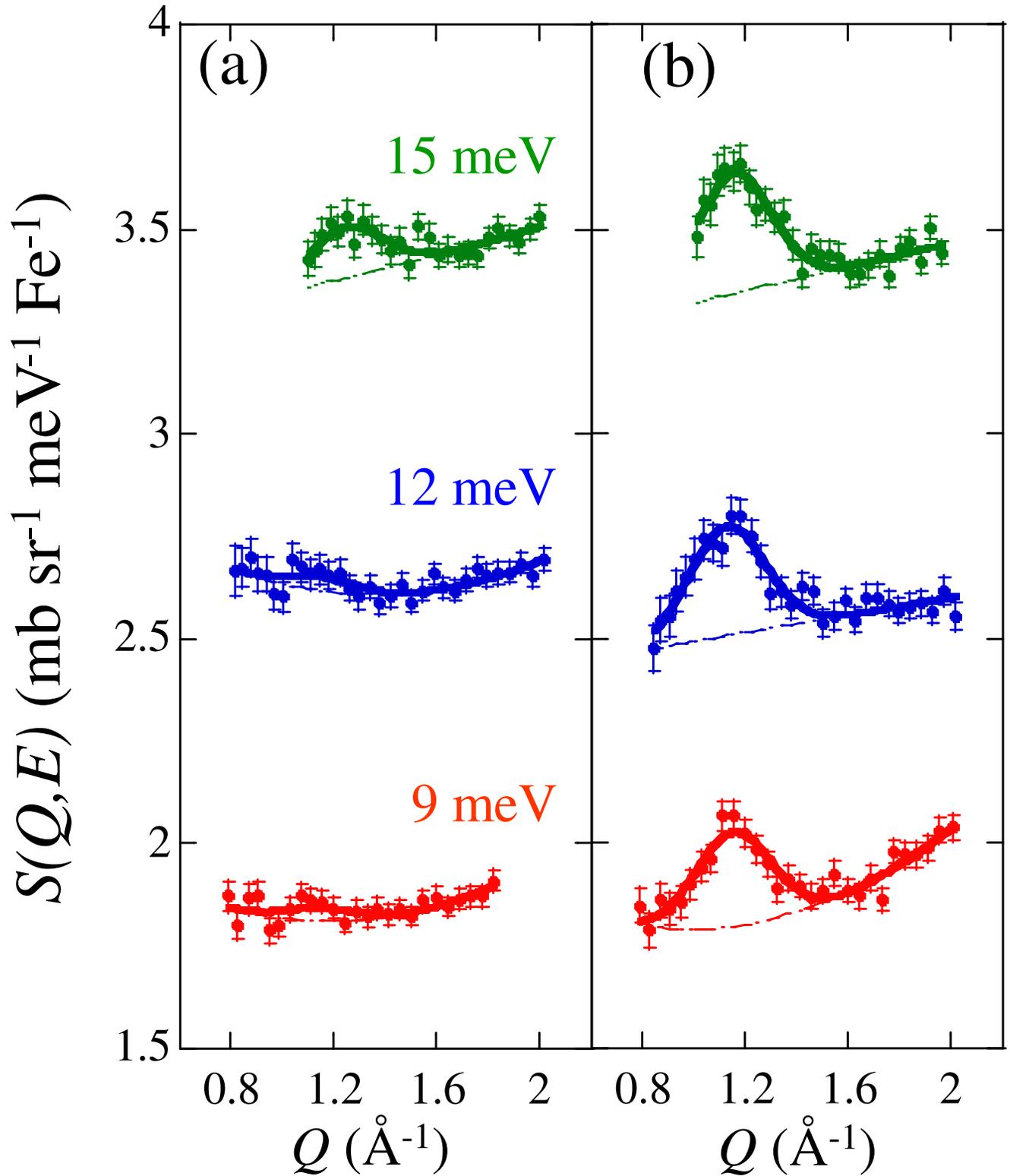}
\caption{Constant-$E$ cuts of the dynamical structure factor $S(Q,E)$ at $T$= 30 K for (a)  LaFePO$_{0.9}$ and (b) LaFeAsO$_{0.918}$F$_{0.082}$. The peak width, full width at half maximum (FWHM), for each fit is fixed to be 0.35 ${\rm \AA^{-1}}$. The $Q$-resolution at $Q$ $\sim$ 1.1 \AA$^{-1}$ is less than 0.12 \AA$^{-1}$, which is much smaller than the observed widths. Scattering patterns are vertically shifted for clarity.}
\label{Figure2}
\end{figure}

The constant-energy ($E$) cuts of the dynamical structure factor $S(Q,E)$ at each momentum and energy transfer ($Q$ and $E$) for LaFePO$_{0.9}$ and the optimally doped LaFeAsOF at 30 K are shown in Fig. \ref{Figure2}. The low-$Q$ region is limited by a kinematic condition with the incident and final wave vectors. For low-energies ranging from 9 to 15 meV, the spin fluctuation is clearly observed only for the optimally doped LaFeAsOF as a peak at about 1.1 ${\rm \AA}^{-1}$, corresponding to the $Q$ = ($\pi$, 0) position in the reduced tetragonal unit cell with $a \sim$ 2.8 ${\rm \AA}$ where the AF fluctuation is commonly observed \cite{Ishikado, Rahn, Wakimoto}. The 2-dimensional spin fluctuation is expected to appear as a magnetic rod in the $Q$-space. The magnetic rod intensity rapidly decreases with increasing $Q$ due to the magnetic form factor of Fe$^{2+}$. Although the magnetic rod signal has a small tail at a higher $Q$ position after averaging the Debye ring, the signal can be approximated as a single Gaussian peak due to the broad width. For LaFePO$_{0.9}$, the peak is strongly suppressed compared to that of LaFeAsO$_{0.918}$F$_{0.082}$. The magnetic resonance mode is expected to be at 2.4 meV in LaFePO$_{0.9}$ based on the simple linear dependence of the energy on $T_{\rm c}$ \cite{Paglione, J.T.Park, Shamoto10}.

Even at energies below 3 meV, spin fluctuations and the magnetic resonance mode in LaFePO$_{0.9}$ were not observed in Disk chopper time-of-flight spectrometer, IN5, and Cold Neutron Triple-Axis Spectrometer, CTAX, (See the Supplementary information). These results are consistent with a previous report for LaFePO \cite{Taylor}.

\begin{figure}[ht]
\centering
\includegraphics[width=\linewidth]{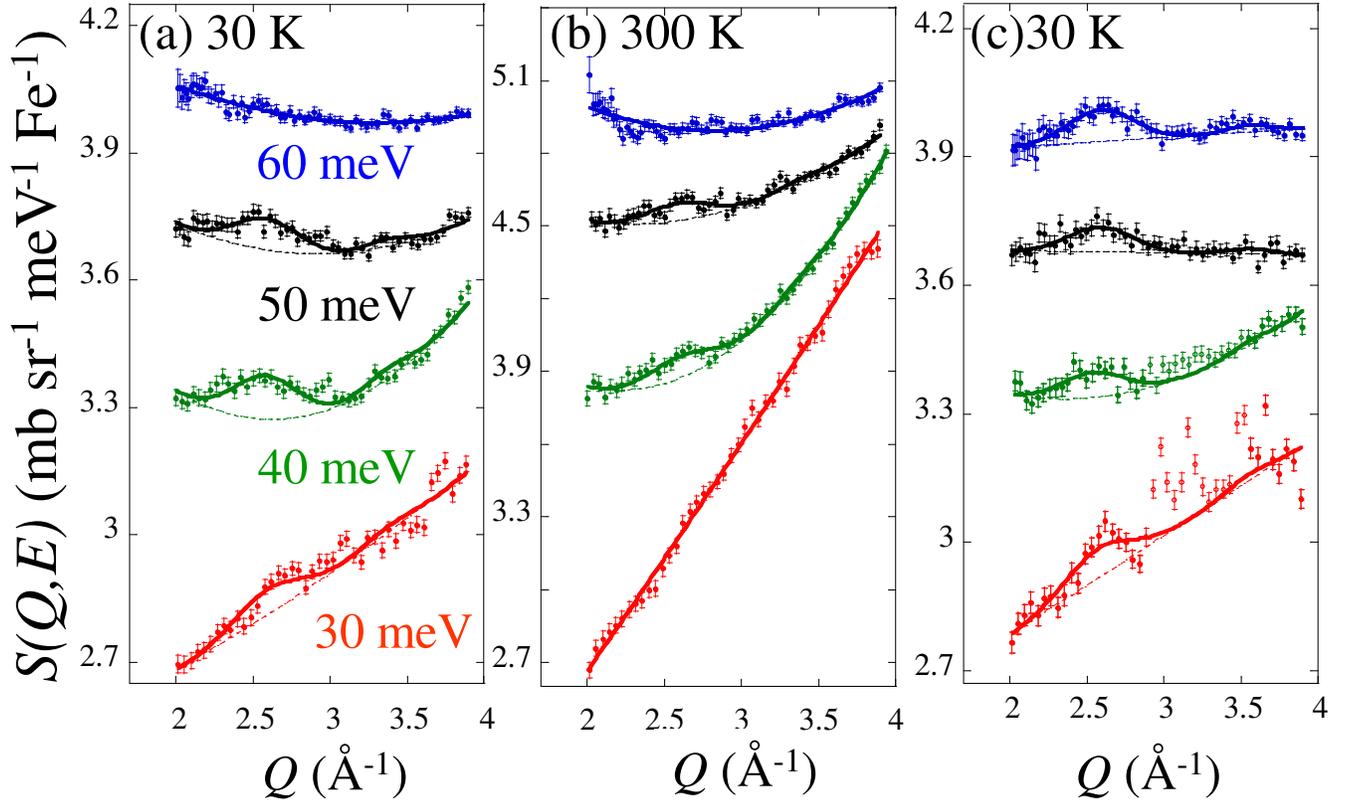}
\caption{Constant-$E$ cuts of the dynamical structure factor $S(Q,E)$ in the energy range from 30 to 60 meV for (a) LaFePO$_{0.9}$ at $T$= 30 K, (b)  LaFePO$_{0.9}$ at $T$= 300 K, and (c) LaFeAsO$_{0.918}$F$_{0.082}$ at $T$= 30 K. The solid lines are fits with fixed intensity ratios. The FWHMs are fixed to be 0.5 ${\rm \AA^{-1}}$ at ($\pi$, 2$\pi$) and 0.35 ${\rm \AA^{-1}}$ at (3$\pi$, 0). The latter width is fixed based on the width at ($\pi$, 0). The $Q$-resolution at $Q$ $\sim$ 2.6 \AA$^{-1}$ is less than 0.25 \AA$^{-1}$ and that at $Q$ $\sim$ 3.4 \AA$^{-1}$ is less than 0.18 \AA$^{-1}$, which is much smaller than the observed widths. Scattering patterns are vertically shifted for clarity.}
\label{Figure3}
\end{figure}

At high energies, however, a spin fluctuation in LaFePO$_{0.9}$ was found as shown in Fig. \ref{Figure3}. One may doubt that the signal may be originated by phonons. However, the difference between phonon and magnetic signals can be distinguished by the $Q$ and $T$ dependences of the intensity. For the $Q$ dependence, 2-dimensional spin fluctuations are expected to appear as magnetic rods at various $Q$ positions such as ($\pi$, 0), ($\pi$, 2$\pi$), (3$\pi$, 0), (3$\pi$, 2$\pi$), etc., due to the periodicity of Brillouin zone\cite{Rahn}. As shown in Fig. \ref{Figure3}, the first peak of $Q$ = ($\pi$, 2$\pi$) appeared at $Q$=2.6 ${\rm \AA^{-1}}$, whereas the second peak of $Q$ = (3$\pi$, 0) appeared at $Q$=3.4  ${\rm \AA^{-1}}$. The intensity depends on the multiplicity and the magnetic form factor of Fe$^{2+}$. In the case of LaFePO$_{0.9}$, the intensity ratio of ($\pi$, 2$\pi$) to (3$\pi$, 0) becomes about 4. The expected small peaks were observed at around 3.4 ${\rm \AA^{-1}}$ in Fig.\ref{Figure3} including LaFeAsO$_{0.918}$F$_{0.082}$. The intensity ratio and the peak width were fixed in the fits in Fig. \ref{Figure3}. In addition, the magnetic inelastic signals are usually observed along the transverse directions (e.g. $K$ direction at (1,0)) to the in-plane scattering vector \cite{Rahn}. This has been attributed to the strong inter-band scattering along the longitudinal direction \cite{Kaneshita}. Because of the present powder data averaging along the Debye ring, peak broadening due to the dispersion was mainly observed at $Q$= 2.6 ${\rm \AA^{-1}}$ in Fig.\ref{Figure3} \cite{Rahn}. The fitted intensities of the constant-$E$ cuts of Fig. \ref{Figure2} at $Q$ $\sim$ 1.1 ${\rm \AA^{-1}}$, and Fig. \ref{Figure3}(a), (c) at $Q$ $\sim$ 2.6 ${\rm \AA^{-1}}$ were converted to $\chi$"$(E)$ by using Bose factor, multiplicity, and magnetic form factor of Fe$^{2+}$, as shown in Fig. \ref{Figure4}.

For the temperature dependence, magnetic signals usually become weak at high temperatures mainly due to the shortening of lifetime. On the other hand, phonon intensity simply increases with increasing temperature based on the Bose factor. Figure \ref{Figure3} shows the spin fluctuations in LaFePO$_{0.9}$ measured at $T$= 30 and 300 K. Based on the Bose factors in the $E$-range, the intensities at $T$= 300 K should increase by factors ranging from 1.11 to 1.46 compared to those at $T$= 30 K. Contrarily, all the intensities decreased with increasing temperature by factors ranging from 0.00 to 0.82. In addition to the two peaks in the INS pattern, this opposite temperature dependence strongly supports that the observed signals are originated from spin fluctuations.

\section*{Discussion}

Any strong AF spin fluctuations have not been observed previously for both LaFePO$_{0.9}$ and the optimally doped LaFeAsOF in the nuclear spin relaxation rate 1/$T_{1}T$ of nuclear magnetic resonance (NMR) measurements \cite{Nakai, Mukuda, Nakai09}. Here, by using INS, they were observed at high energies. For LaFePO$_{0.9}$, there is a large energy gap greater than 12 meV in the $\chi$"$(E)$. This energy dependence is consistent with a theoretical calculation by DFT+DMFT \cite{Yin}, in addition to there being no enhancement of the nuclear spin relaxation rate 1/$T_{1}T$ of NMR measurements. This can also be explained by the wide effective band width of the magnetic excitations from the relatively low pnictogen height in LaFePO$_{0.9}$. The observed intensity of $\chi$"$(E)$ in LaFePO$_{0.9}$ at 30-50 meV is greater than that of the optimally doped LaFeAsO$_{0.918}$F$_{0.082}$. These strong spin fluctuations are consistent with line-node symmetry in spin-fluctuation-mediated superconductors \cite{Fletcher, Yamashita}. 

\begin{figure}[ht]
\centering
\includegraphics[width=\linewidth]{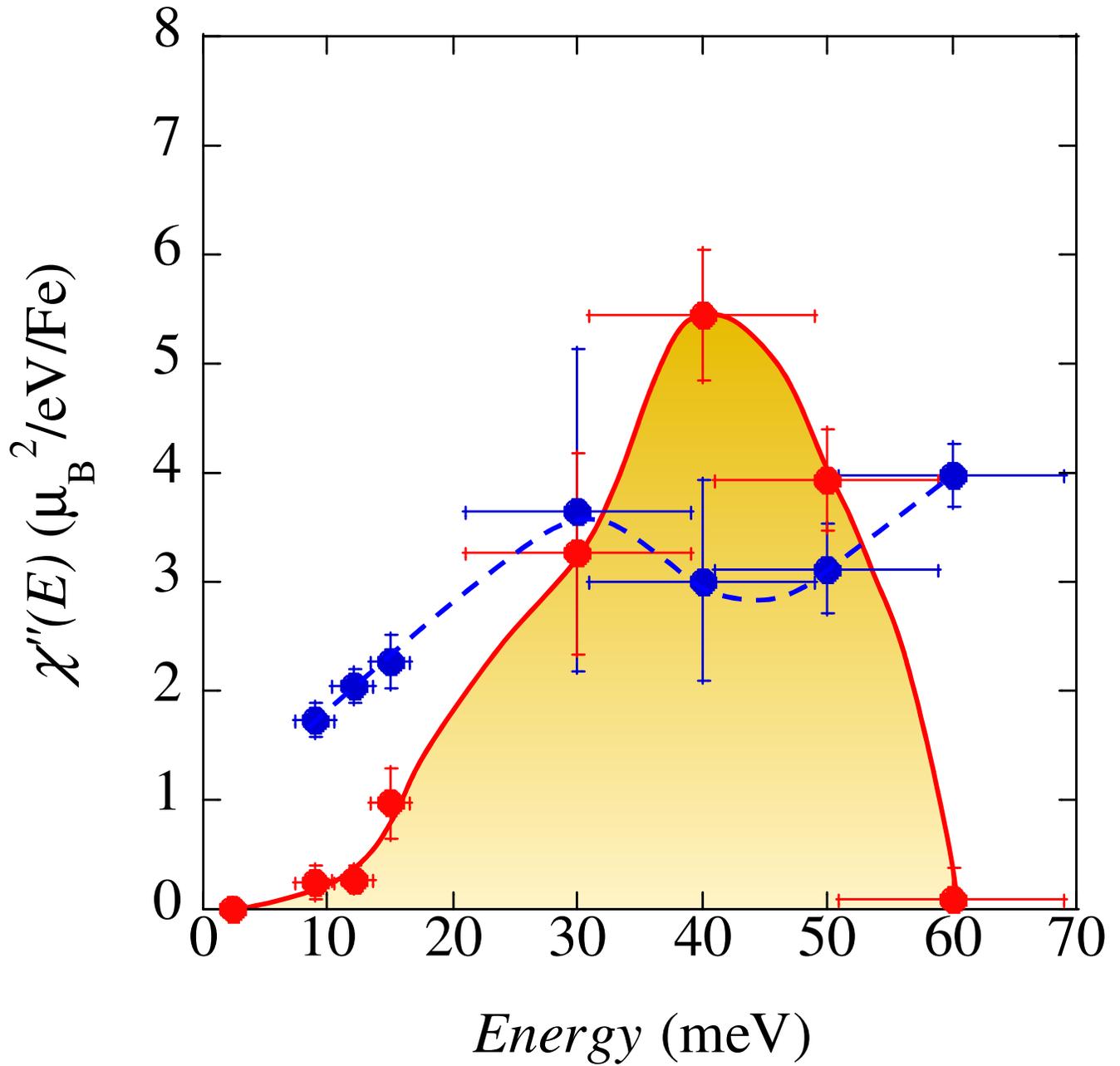}
\caption{Momentum-integrated dynamical spin susceptibility $\chi$"$(E)$ for LaFePO$_{0.9}$ (red filled circles) and LaFeAsO$_{0.918}$F$_{0.082}$ (blue filled circles). All the points were measured at the normal state of $T$= 30 K. The solid and broken lines are guides for the eye.}
\label{Figure4}
\end{figure}

On the other hand, the spin fluctuations in the optimally doped LaFeAsOF have a much lower energy component, as shown in Fig. \ref{Figure4}. It is intriguing to point out that the absolute value of $\chi$"$(E)$ at about 20 meV ($\sim$3 $\mu_{\rm B}^{2} {\rm eV}^{-1}{\rm Fe}^{-1}$) for LaFeAsOF is also very similar to the value $\sim$3 $\mu_{\rm B}^{2} {\rm eV}^{-1} {\rm Cu}^{-1}$ of La$_{1.84}$Sr$_{0.16}$CuO$_{4}$\cite{Vignolle} (in addition to having similar $T_c$ value) although their electronic structures, i.e., single and multi-orbitals, are very different. This similarity can be a hint for these unconventional superconductivity from the spin fluctuation point of view. 

The existence of low-energy spin fluctuations may be related to the proximity to the AF parent compound with a structural phase transition. For the LaFeAsOF system, the low-temperature enhancement of 1/$T_{1}T$ is observed only in the vicinity of the AF ordered phase. Upon electron doping, the enhancement is rapidly suppressed, suggesting an energy gap in $\chi$"$(E)$ at low energies \cite{Nakai09}. The AF spin fluctuations, however, are found here to persist at high energies in this work. The low-$E$ spin fluctuation below 10 meV seems to disappear depending on the material parameter, $U$/$W$ (electron-electron correlation energy, $U$, and electronic band width, $W$) by increasing the doping or $W$. The detailed structure of $\chi$"$(E)$ has some discrepancies from DFT+DMFT calculation. For example, the observed peak structure at 30-50 meV in LaFePO$_{0.9}$ (Fig. \ref{Figure4}) does not appear in the calculation. In addition, according to the calculation, the intensity for LaFePO is one order of magnitude smaller than that of LaFeAsO. Therefore, it is necessary to study in details how the energy dependence of the spin fluctuations depends on the material parameters experimentally . 

\section*{Methods}

Polycrystalline samples of LaFePO$_{0.9}$, and LaFeAsO$_{0.918}$F$_{0.082}$ were prepared by a solid-state reaction method. LaP, LaAs, Fe$_2$O$_3$, Fe, and FeF$_2$ powders were used as the starting materials. LaP(As) was obtained by reacting La powders and P (As) grains in an evacuated quartz tube at 500${}^\circ\mathrm{C}$ for 5 h and then 700${}^\circ\mathrm{C}$ (850${}^\circ\mathrm{C}$) for 10 h. The starting materials were ground with the nominal compositions LaFePO$_{0.9}$ and LaFeAs(O$_{0.9}$F$_{0.1}$)$_{0.9}$ using agate mortar and then were pressed into pellets. Note that the molar ratio of (O$_{1-x}$F$_x$)$_{0.9}$ is lower than the stoichiometry because of partial oxidation of the precursor. They were then sintered for 10 h in an evacuated quartz tube at a sintering temperature 1250${}^\circ\mathrm{C}$ for LaFePO$_{0.9}$ and 1100${}^\circ\mathrm{C}$ for LaFeAsO$_{1-x}$F$_x$. The heating rate was kept below 50${}^\circ\mathrm{C}$/h to prevent the explosion of the quartz tube due to the sudden increase of the P(As) vapor pressure. All the processes were performed in a glove box filled with nitrogen or helium gas. The actual fluorine content of LaFeAsO$_{1-x}$F$_x$ was determined by Secondary Ion Mass Spectrometry \cite{Wakimoto} to be 0.082. X-ray diffraction patterns were measured using Cu K$\alpha$ radiation (Rigaku RINT 1100), and the observed peaks were indexed to the tetragonal ZrCuSiAs-type (so-called 1111-type) structure with space group of $P4/nmm$ and the lattice parameters were $a$ = 3.955 ${\rm \AA}$ and $c$ = 8.504 ${\rm \AA}$ for LaFePO$_{0.9}$ and $a$ = 4.026 ${\rm \AA}$ and $c$ = 8.724 ${\rm \AA}$ for LaFeAsO$_{0.918}$F$_{0.082}$ as shown in Fig. 1(b). The dc magnetic susceptibility was measured using a SQUID magnetometer (MPMS, Quantum Design Inc.) under a magnetic field of 5 Oe. As shown in Fig. 1 (a), the $T_c$ values were determined to be 5 K for LaFePO$_{0.9}$ and 29 K for LaFeAsO$_{0.918}$F$_{0.082}$. Among three samples of LaFePO$_{1-y}$ with different initial oxygen contents ($y$=0.0, 0.1, 0.2), the largest superconducting shielding volume fraction of $\sim$100\% was observed for $y$=0.1. On the other hand, the stoichiometric compound, LaFePO$_{1-y}$ ($y$=0.0) showed no superconductivity, which is consistent with a previous report \cite{McQueen}. 

INS measurements were carried out using three spectrometers for high- and low- energy region complementally. High-energy measurements were performed using a Fermi chopper spectrometer at 4D Space Access Neutron Spectrometer (4SEASONS), BL01, in Japan Proton Accelerator Research Complex (J-PARC) Materials and Life Science Experimental Facility (MLF). The incident energies of $E_i$ =45.5 and 150 meV were employed with the multi-$E_i$ method \cite{Kajimoto, Nakamura}. The energy resolutions are 3.2 and 18.0 meV for $E_i$ =45.5 and 150 meV at around $E$=0 meV, respectively. These values were obtained by estimating the full width at half maximum (FWHM) of the incoherent scattering around zero-neutron energy transfer. The measurement time and the sample weight are 11.5 h - and 25 g, respectively, for LaFePO$_{0.9}$ and 22 h - and 25 g, respectively, for LaFeAsO$_{0.918}$F$_{0.082}$ at a beam power of 280 kW. To access the low-$E$ region corresponding to the expected magnetic resonance energy of about 2 meV of LaFePO$_{0.9}$, a cold triple-axis spectrometer (CTAX) at High Flux Isotope Reactor (HFIR) in ORNL and chopper spectrometer IN5 in ILL were used. The measured transferred energy ranged from 1.0 to 3.0 meV from the CTAX and the neutron wavelength was $\lambda$=4.5 ${\rm \AA}$ at IN5. The same 34 g sample was split into 14 g and 20 g for each spectrometer. For both measurements, orange cryostats were used to access low temperatures down to 1.5 K.

The constant-$E$ plot was fit by Gaussian functions with constant, linear and quadratic functions as backgrounds. The absolute value of dynamical structure factor, $S(Q,E)$, is normalized by the Bragg peak intensity at (002). We also confirmed that the absolute value of the $S(Q,E)$ coincided within 10\% by using another method of vanadium normalization. The dynamical spin susceptibility is estimated as an isotropic spin fluctuation \cite{Hayden}. {\sc Utsusemi} software was used for data analysis of the data sets obtained at 4SEASONS \cite{Inamura}.


\section*{Acknowledgements}

The authors thanks to Prof. M. Arai, Drs. Y. Inamura, S. Wakimoto, A. Iyo, H. Eisaki, F. Esaka and Mr. F. Mizuno for their helpful discussions. Magnetic susceptibility measurements were performed by using the SQUID magnetometer (MPMS,Quantum Design Inc.) at the CROSS user laboratory (B402). INS experiments at J-PARC MLF were carried out under project numbers 2009A0087, 2014A0114 and 2017I0001. The experiment at ILL was conducted under the experiment number DIR-97 as a support program (Director's Discretion Time (DIR)) to the earthquake damage of Japanese neutron facilities. This work was supported by JST, Transformative Research-Project on Iron Pnictides (TRIP), Grant-in-Aid for Specially Promoted Research, Ministry of Education, Culture, Sports, Science and Technology (MEXT), Japan (No. 17001001), and JSPS KAKENHI Grant Number JP15K17712. A portion of this research used resources at the High Flux Isotope Reactor, a DOE Office of Science User Facility operated by the Oak Ridge National Laboratory, and was partly supported by the US-Japan Collaborative Program on Neutron Scattering.\\

\section*{Author contributions statement}

M.I., S.S., K.K., R.K., M.N., T.H. and H.M. conducted the inelastic neutron scattering measurements. M.I. synthesized and characterized the powder samples. S.S. designed and coordinated the experiments. S.S. and M.I. mainly contributed to the manuscript with help from the others. All authors reviewed the manuscript and the comments were taken into consideration.

\section*{Additional information}

Competing financial interests: The authors declare no competing interests.

\end{document}